# Anomalous conductivity dependence of plasticized PVC for different modificator "A" concentrations and film thicknesses

D. V. Vlasov, L. A. Apresyan, A. D. Vlasov and V. I. Kryshtob

Prokhorov General Physics Institute, Russian Academy of Science

ul.Vavilova 38, Moscow, 119991 Russia

## ABSTRACT

**The dependences of electrical conductivity of plasticized PVC films on mass fraction of plasticizer "A" and the film thickness are experimentally investigated. Non-monotonic dependence of conductivity on the concentration of plasticizer and strongly nonlinear dependence of the resistance of the film on its thickness are found. Possibility of construction of the models describing received results is shown and also discussed.**

**Аномальная зависимость проводимости поливинилхлоридных пленок от их толщины и массовой доли пластификатора «А»**

**Введение**

В процессе исследования электрофизических характеристик пластифицированных модификатором «А» [1,2] пленок ПВХ [3-9] толщиной более 20 мкм, были обнаружены аномальные эффекты во многом сходные с результатами, полученными ранее в работах для субмикронных пленок (см., напр., [10,11]). В отличие от большинства известных авторам работ по изучению электропроводности и, в частности, эффектов переключения в тонких полимерных пленках, в которых использовались пленки с фиксированным составом полимера, в данном сообщении исследуется зависимость проводимости от двух параметров, а именно, концентрации пластификатора (модификатора «А») и толщины пленки.

Для субмикронных пленок из опубликованных ранее результатов известно существование зависящей от природы полимера критической толщины (имеющей порядок одного или нескольких микрон [12] ), при превышении которой аномальные эффекты типа переключения в состояние высокой проводимости (СВП) не наблюдается. В отличие от этого, в рассматриваемом ниже случае пластифицированного ПВХ эффекты переключений сохраняются для значительно больших толщин, вплоть до десятков и сотен микрон. При этом исследование зависимостей от толщины при фиксированных прочих параметрах задачи приобретает особый интерес, давая полезную дополнительную информацию о возможных физических механизмах нелинейной проводимости. Так, в случае субмикронных пленок полидифениленфталида в работе [13] наблюдалась зависимость проводимости от толщины пленки L вида $L^{-3}$, что послужило существенным аргументом в пользу модели инжекционных токов, ограниченных объемным зарядом. В более ранней работе [14] для пленок лавсана, превышающих некоторую характерную толщину, при давлении несколько килобар была обнаружена экспоненциальная зависимость сопротивления от квадрата толщины $L^2$, которая хорошо описывалась предложенной в [14] моделью стимулированной давлением инжекции носителей тока из металла в зону проводимости. В рассматриваемом ниже случае для относительно толстых пленок пластифицированного ПВХ обнаружена экспоненциальная зависимость сопротивления пленки от L (а не от $L^2$), что практически исключает для рассматриваемой области толщин влияние на проводимость инжекционных токов и требует привлечения другой физической модели проводимости.

Далее в настоящей работе описаны результаты экспериментальных исследований зависимостей электропроводности от толщины пленки и содержания пластификатора, и обсуждается возможность построения моделей, описывающих совокупность полученных результатов.

**Эксперимент**

В развитие результатов работ [3-9] по изучению аномалий электропроводности в пластифицированных пленках ПВХ в данной работе проведены измерения проводимости (тока через пленку) для набора пленок одинаковой толщины 20 мкм, но с различным массовым содержанием модификатора "А" ( массовые доли ПВХ: Модификатор «А»== 100:0-100, соответственно). При этом использовалась стандартная схема с последовательным включением всех элементов цепи: балластным сопротивлением ( от 3.3 МОм до 100кОм), собственно стандартной измерительной ячейки с пленкой-образцом пластиката и цифровым считыванием тока посредством считывания напряжения АЦП с прецизионного сопротивления 3К. Для всех образцов с различными массовыми долями пластификатора снимались подробные вольт-амперные характеристики, так что в результате совокупности измерений получились семейства зависимостей тока через пленку пластиката при фиксированном напряжении и изменении в указанных выше пределах массовой доли пластификатора. Анализ зависимостей проводился при напряжениях от нуля до максимального значения 124 В, причем для напряжений, меньших максимального, получаются аналогичные графики с достаточно высокой повторяемостью зависимостей тока от уровня прикладываемого напряжения. Кроме того, все измерения проводились при строго фиксированном давлении, одинаковом для всех типов образцов разной толщины и с различными массовыми долями пластификатора, с использованием стандартной кольцевой ячейки для измерения проводимости полимерных пленок, подробно описанной в [3-9].

Наиболее интересным из полученных результатов можно считать необычную зависимость проводимости пленок пластикатов от массовой доли пластификатора «А». Вместо ожидаемого монотонного увеличения проводимости от уровня типичного изолятора (непластифицированный ПВХ) с величиной удельного объемного сопротивления $\rho = > 10^9$ Ом·см до уровней «антистатики» с $\rho=10^6$ Ом·см, на экспериментальной зависимости проводимости (тока) от массовой доли пластификатора «А» обнаружена аномальная зона (см. Рис. 1), в которой имеет место беспороговый переход в состояние высокой проводимости. При этом сопротивление 20 мкм пленки пластиката в области массовых частей 42.5 – 50 модификатора «А» (на сто массовых частей ПВХ) в исходном состоянии всего около R~ 100 Ом.

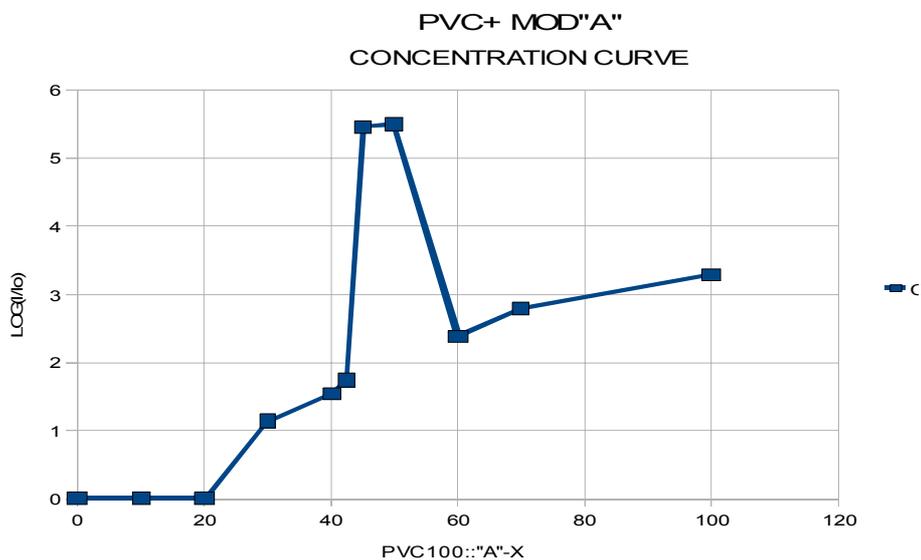

Рис.1 Зависимость тока через пленку пластиката ПВХ толщиной 20 мкм при напряжении 124 В. По оси абсцисс – массовая доля модификатора «А», по оси ординат – десятичный логарифм отношения тока к среднему значению токов утечки (при нулевом внешнем напряжении).

Если условно вычислить (считая, что система линейна и подчиняется закону Ома) удельное сопротивление, то получим $\rho \sim 10^5$ Ом·см, что соответствует уровню нормальных полупроводников, т.е. концентрационно-стимулированный переход в СВП переводит классический изолятор в зону полупроводниковой проводимости. Далее, за исключением этой относительно узкой зоны концентраций, отвечающих беспороговому переходу в СВП, исходное состояние образца соответствует сотням –десяткам ГОм ( в частности, существенно превышает величину балластного сопротивления 3.3 МОм), и как раз попадает в характерные для «антистатика» диапазон удельных сопротивлений.

Таким образом, в соответствии с полученными результатами, на зависимости проводимости полимерной пленки от массовой доли пластификатора можно выделить три зоны:

1. зону «обедненной» массовой доли пластификатора (0-42.5) – в нормальном состоянии проводимость низкая ($R \sim 10^{11}$-$10^{9}$ Ом) ;
2. зону «резонансной» или «нормальной» массовой доли (42.5-50), – в исходном состоянии проводимость высокая ( $R \sim 100$ ОМ)
3. зону с «избыточным» содержанием пластификатора (от 50 и выше) - в исходном состоянии проводимость находится на уровне антистатики ($R \sim 10^{6}$ – $10^{10}$ Ом).

Первоначальные соображения предполагали, что оптимальное для получения антистатических свойств содержание пластификатора в образцах порядка 70-80 м.ч. на 100 м.ч.ПВХ, такие образцы и были исследованы в работах [3-9]. Исходные состояния в этом диапазоне, являются квази-стационарными, т.е. как было обнаружено ранее [3-5], в зоне с «избыточным» содержанием пластификатора (3) наблюдаются спонтанные низковольтные переходы из более стабильного состояния в СВП и обратно. Такие переходы регистрируются экспериментально и применительно к исследованным в данной работе образцам приведены на Рис. 2 и Рис.3.

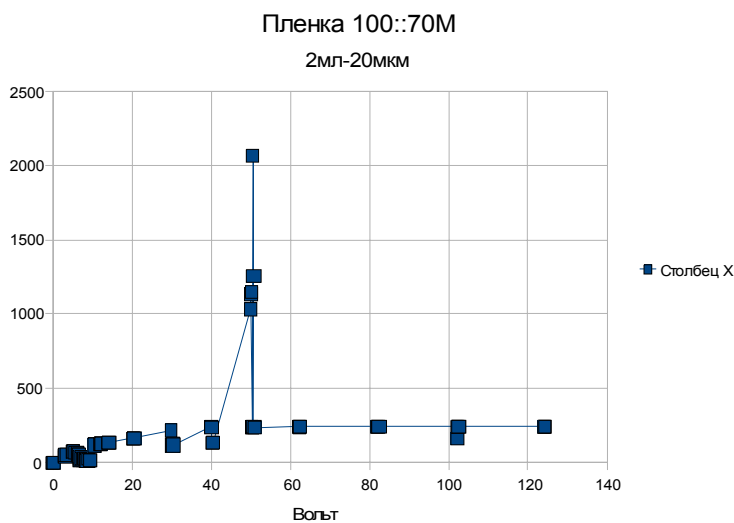

Рис.2. Вольт-амперная характеристика образца пластиката толщиной 20 мкм с массовой долей модификатора «А»- 70 м.ч. со случайным выбросом тока (переход в СВП) в области 40-45 В. По оси абсцисс – Вольты, по оси ординат - значения тока через образец. Скорость снятия характеристики 5 сек на точку (усреднение по 10 отсчетам)

Для кривой на Рис.2, изображающей зависимость тока от прикладываемого напряжения, характерна тенденция к насыщению тока по мере увеличения приложенной разности потенциалов (отмечавшийся ранее эффект стабилизации тока [3-5]). Кроме того, при разных напряжениях (не обязательно максимальных) наблюдаются более или менее продолжительные спонтанные выбросы тока вверх с возвратом в исходное состояние с низкой проводимостью (см. подробнее [3-5]).

В зоне «резонансной» или «нормальной» массовой доли пластификатора имеет место совершенно иная вольтамперная характеристика. В частности, исходное состояние отвечает СВП, поэтому зависимость тока с незначительными отклонениями соответствует закону Ома на балластном сопротивлении (поскольку в измерениях применялись балластные сопротивления порядка МОм, а сопротивление образца в нормальном состоянии составляло сотни Ом).

Следует отметить, что как и в зоне с «избыточным» содержанием пластификатора, исходное состояние высокой проводимости квази-стабильно, однако в этом случае переходы, как это видно из Рис.3, происходят из состояния с высокой проводимостью вниз в состояние с низкой проводимостью (СНВ). Вольтамперная характеристика при этом «соскакивает» с закона Ома и «перескакивает» на значения характерные для кривой на Рис. 2 , но затем вновь возвращается в исходное состояние.

Поскольку измерения вольтамперных характеристик проводилось для каждой точки, можно с уверенностью говорить о некоторой статистике и воспроизводимости свойств, и в частности, переходов как в СВП в зоне с «избыточным» содержанием пластификатора, так и в СНВ в зоне «резонансной» массовой доли пластификатора.

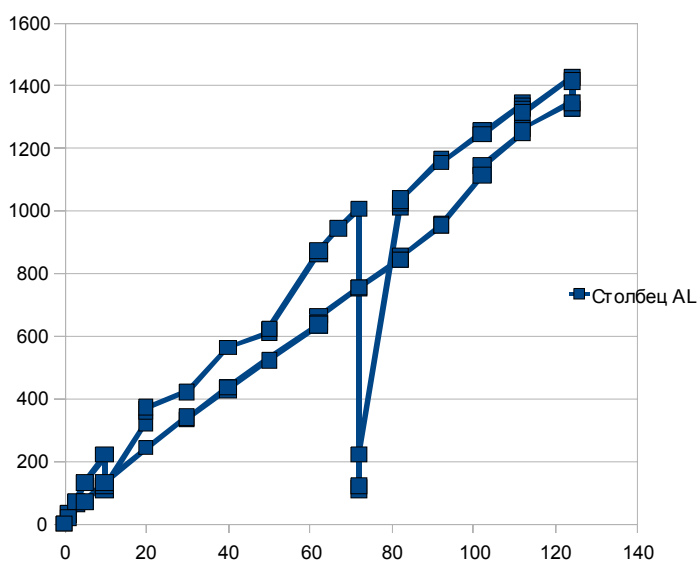

Рис.3 Вольт-амперная характеристика образца пластиката толщиной 20 мкм с массовой долей модификатора «А» -45 м.ч. со случайным выбросом вниз в районе 65 В. По оси абсцисс – Вольты, по оси ординат- значения тока через образец в условных единицах. Скорость снятия характеристики 5 сек на точку (усреднение по 10 отсчетам)

Еще одна часть измерений связана с определением зависимости электропроводности пленки при «резонансной» концентрации от толщины при фиксированных остальных параметрах задачи (доли пластификатора, температуры и давления). Результаты таких измерений показаны на Рис.4 в логарифмическом по проводимости масштабе. Эти графики хорошо аппроксимируются простой экспоненциальной зависимостью от толщины, в отличие от аналогичных результатов для тонких пленок из других полимеров, о которых говорилось во Введении.

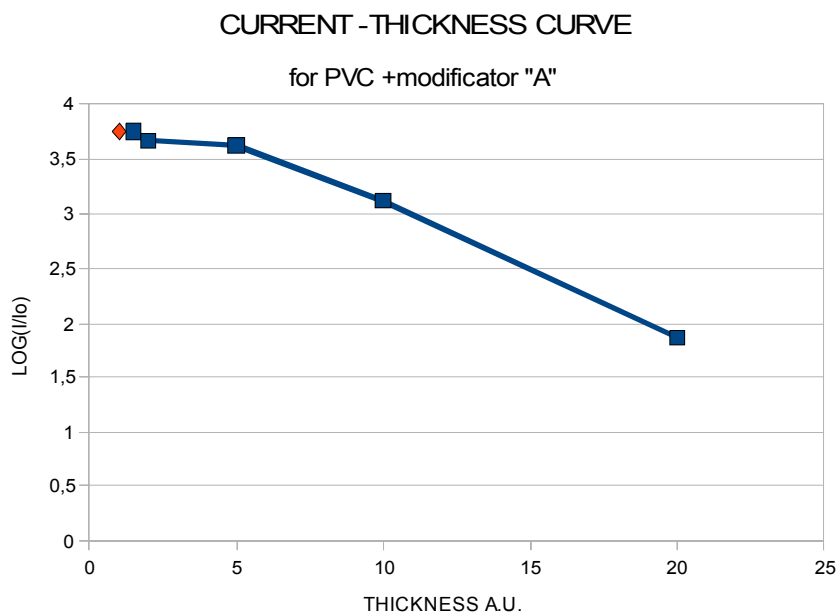

Рис.4. Экспериментальная зависимость тока от толщина образца пластиката. По оси абсцис толщина образца в произвольных единицах близких к мкм,- по оси ординат десятичный логарифм тока.

Из Рис. 4 следует, в частности, что при прочих равных условиях, ток через образец пластиката при увеличении его толщины падает по экспоненциальному закону, что согласуется формулой определяющей ток в рамках развиваемой модели. Отметим, что ранее соответствие модели и экспериментальных результатов обсуждалось на качественном уровне, фактически приведенные выше формулы и функциональное сопоставление с экспериментальными данными получено впервые.

**Обсуждение результатов и выводы**

Наличие скачка проводимости на четыре-пять порядков при плавном и монотонном увеличении массовой доли пластификатора (Рис.1) на наш взгляд свидетельствует о сложной и локально неоднородной внутренней структурной организации пленок пластикатов, приготовленных методом полива. Первая фаза и начало второй в некоторой степени напоминает график изменения проводимости при прохождении порога перколяции при наполнении изолирующей матрицы проводящими включениями [15]. Максимально упрощая модель, можно сказать, что порог перколяции или резкий скачок проводимости композита возникает, когда происходит формирование бесконечного кластера и связанного с ним проводящего канала из последовательно соприкасающихся случайным образом проводящих частиц в изолирующей матрице. Однако, в отличие от композитов с проводящими включениями, в рассматриваемом случае в непроводящую диэлектрическую матрицу добавляется второй компонент - также непроводящий в исходном состоянии пластификатор, который при определенной концентрации приводит к скачку проводимости. Поэтому обычная теория протекания не может быть использована для объяснения скачка проводимости, наблюдаемого с увеличением доли пластификатора «А». С другой стороны, приведенные выше результаты, как уже отмечалось во Введении, вряд ли могут быть объяснены и на основе теорий инжекции носителей тока из металлических электродов [13,14].

       Наиболее разумное, по мнению авторов, предположение, которое можно сделать на основании полученных данных — это высокая проводимость возникающая на границах двух взаимопроникающих изоляторов: ПВХ и пластификатора «А» (отметим, что примеры наблюдения аномального роста проводимости на макроскопической границе двух изоляторов и, в частности, аморфных полимеров наблюдаются и активно исследуется в настоящее время [16]). Применительно к кластерам ПВХ, связанным с пластификатором, в [5] была предложена качественная молекулярно-электронная модель проводимости пластиката. В этой модели заряд переносится ставшими подвижными вследствие присутствия пластификатора фрагментами (кинетическими сегментами) макромолекул ПВХ (на расстояния, не превышающие длину подвижного сегмента).

       В рамках развиваемой модели в процессе формирования получаемой методом полива пленки пластиката происходит стратификация истинного (менее 1%) молекулярного раствора макромолекул ПВХ и пластификатора (в соответствующей массовой доле) на кластеры молекул ПВХ и связанные с ними Ван-дер-Ваальсовыми взаимодействиями молекулы пластификатора. При этом макромолекулы ПВХ и находящиеся с ними в контакте молекулы пластификатора участвуют в процессе переноса заряда. Так за счет пластификатора возникают подвижные сегменты в молекулах ПВХ, с помощью которых и осуществляется последовательный перенос зарядов, которые как бы скользят по поверхности ПВХ кластеров в направлении, соответствующем приложенному электрическому полю. Такая модель переноса зарядов фактически эквивалентна развиваемой ранее авторами [5-9], однако с весьма существенным дополнением о том, что проводимость ПВХ кластеров существует на его поверхности,

соприкасающейся с оболочкой пластификатора. Таким образом, обычный прыжковый механизм переноса зарядов дополняется учетом теплового движения образуемых подвижными сегментами ловушек.

Принимая эту модель можно фактически получить ситуацию, эквивалентную порогу перколяции при анализе системы ПВХ-пластификатор, причем «замыкание канала проводимости» между электродами должно осуществляться через цепочку поверхностей кластеров ПВХ, соприкасающихся с пластификатором. Скачкообразный рост проводимости при плавном увеличении концентрации пластификатора можно описать моделью, аналогичной теории протекания, однако в отличие от обычной перколяции, дальнейшее увеличение массовой доли пластификатора (в нашем случае до примерно 50%) приводит к обратному переключению в начальное состояние с низкой проводимостью. Этот второй переход — скачок проводимости вниз, в рамках рассматриваемой модели находит логическое объяснение также из общих соображений,- поскольку оба компонента смеси изоляторы, избыточная концентрация любого компонента должна приводить к диэлектрическому уровню проводимости. Для конкретной модели пластикатов ПВХ, где перенос заряда осуществляется подвижными сегментами макромолекул, дистанция переноса заряда ограничивается средним размером сегмента, и в том случае, когда толщина прослоек пластификатора превышает этот размер сегмента - канал проводимости «обрывается» и результирующее сопротивление растет скачкообразно. Отметим, что низковольтные спонтанные переключения проводимости в СВП и обратно наблюдавшиеся ранее [3-5], очевидным образом объясняются в рамках развиваемой модели, которую можно было бы называть объемно-интерповерхностной проводимостью поскольку микроскопические подвижки кластеров ПВХ могут приводить к восстановлению или разрыву канала.

Весомым подтверждением рассматриваемой модели проводимости является необычная зависимость сопротивления от толщины пленки. В работе [9] была впервые получена формула для зависимости сопротивления плёнки от её толщины. Мы повторим этот вывод здесь: поскольку в описываемых исследованиях эта формула получила экспериментальное подтверждение. Обозначим $f(L)$ количество проводящих каналов, проходящих через всю толщину плёнки. Естественно предположить, что эти каналы независимы и места выхода этих каналов на поверхность случайны. Мысленно разобьём рассматриваемую плёнку на две, толщинами $L_1$ и $L_2$, $L_1 + L_2 = L$. Тогда число каналов на единицу площади в них будет равно, соответственно $f(L_1)$ и $f(L_2)$. Но не по всем каналам ток сможет протечь через обе плёнки, а только по тем, которые проходят по всей толщине полимера. Другими словами, ток может пройти только через те каналы, которые встречаются на границе раздела между плёнками толщинами $L_1$ и $L_2$. Посчитаем число таких каналов. Пусть общая площадь рассматриваемого участка равна S, а площадь поверхности одного канала σ. Тогда число каналов в одной плёнке равно $f(L_1)S$, в другой $f(L_2)S$. Посчитаем вероятность того, что какой-то выделенный канал с верхней плёнки встретится с каким-то каналом из нижней плёнки. Для этого канал нижней плёнки должен попасть в зону вокруг канала верхней плёнки, площадь этой зоны 4σ. Среднее число каналов (точнее, математическое ожидание числа каналов) нижней плёнки на площадке 4σ равно $f(L_2) 4σ$. Всего таких площадок столько же, сколько и каналов верхней плёнки, т. е. $f(L_1)S$. Таким образом, всего каналов, которые встречаются сверху и снизу, получается $f(L_1)S f(L_2) 4σ$, и это равно $f(L)S$. Таким образом, получаем функциональное уравнение для $f(L)$:

$$f(L_1+L_2) = 4σ\, f(L_2)\, f(L_1)$$

Это уравнение имеет решение:

$$f(L) = \frac{1}{4σ} e^{-\frac{L}{L_0}}$$

Хочется обратить внимание, что предэкспоненциальный множитель определяется из функционального уравнения однозначно, а $L_0$ — произвольный параметр размерности длины. Эта формула была получена в весьма грубых предположениях — например, каналы могут пересекаться почти по всей площади, а могут только чуть-чуть касаться друг друга, а эти случаи считались одинаковыми. Это отразится на том, что перед экспонентой будет не 4σ, а σ с каким-то другим коэффициентом. Здесь мы также пренебрегли краевыми эффектами (т. е. тем, что около реальной границы полимера каналы могут вести себя не так, как внутри), тем, что распределение каналов может не быть независимым, что разные слои могут не быть независимыми, что каналы могут быть разных размеров и т. д.

Отметим, что аномальные результаты зависимости тока образцов пластикатов, описанные в настоящей работе достаточно убедительно коррелируют с экспериментальными результатами по исследованию аномалий проводимости пластикатов ПВХ, полученными авторами ранее [3-9], а в сочетании с подробной и подтверждаемой экспериментально моделью, превращает «аномальные» результаты в логически связанные естественные зависимости. Тем не менее, пока модель и интерпретация экспериментальных данных носит в основном качественный характер.

Заметим также, что развитая модель может быть применена, по-видимому, для всех полимерных систем, в которых реализуются каналы проводимости (хорошо известные в литературе [3-15]). Отчасти наблюдаемая экспоненциальная зависимость объясняет общее стремление работать с тонкими пленками, в более толстых пленках за счет экспоненциальной зависимости ток зарегистрировать представляется крайне затруднительным. Другим важным следствием полученных результатов является отмечавшаяся ранее существенная ограниченность использования понятия удельной объемной проводимости полимерных материалов с проводимостью канального типа.

Таким образом, наблюдаемые в пластифицированных модификатором «А» пленках ПВХ аномальные зависимости электропроводности от содержания пластификатора и толщины пленки допускают качественное, и, отчасти, количественное объяснение в рамках модели «подвижных ловушек», предложенной ранее в [5]. Можно ожидать, что указанная модель при дальнейшем развитии будет полезной для описания аномалий электропроводности и в случае пленок из других полимеров, для которых подвижность сегментов может регулироваться количественно введением пластификаторов.

**Литература**